# Basin Effects in Strong Ground Motion:
# A Case Study from the 2015 Gorkha, Nepal Earthquake

Peyman Ayoubi,[1] Domniki Asimaki, Ph.D.,[2]
and Kami Mohammdi, Ph.D.[3]

[1]Department of Mechanical and Civil Engineering, California Institute of Technology, 1200 E California Blvd, Pasadena, Ca; e-mail: ayoubi@caltech.edu
[2]Department of Mechanical and Civil Engineering, California Institute of Technology, 1200 E California Blvd, Pasadena, Ca; e-mail: domniki@caltech.edu
[3]Department of Mechanical and Civil Engineering, California Institute of Technology, 1200 E California Blvd, Pasadena, Ca; e-mail: kamimohamadi@caltech.edu

## ABSTRACT

The term "basin effects" refers to entrapment and reverberation of earthquake waves in soft sedimentary deposits underlain by concave basement rock structures. Basin effects can significantly affect the amplitude, frequency and duration of strong ground motion, while the cone-like geometry of the basin edges gives rise to large amplitude surface waves through seismic wave diffraction and energy focusing, a well-known characteristic of basin effects. In this research, we study the role of basin effects in the mainshock ground motion data recorded at the Kathmandu basin, Nepal during the 2015 $M_w$7.8 Gorkha earthquake sequence. We specifically try to understand the source of the unusual low frequency reverberating pulse that appeared systematically across the basin, and the unexpected depletion of the ground surface motions from high frequency components, especially away from the basin edges. In order to do that we study the response of a 2D cross section of Kathmandu basin subjected to vertically propagating plane SV waves. Despite the scarcity of geotechnical information and of strong ground motion recordings, we show that an idealized plane-strain elastic model with a simplified layered velocity structure can capture surprisingly well the low frequency components of the basin ground response. We finally couple the 2D elastic simulation with a 1D nonlinear analysis of the shallow basin sediments. The 1D nonlinear approximation shows improved performance over a larger frequency range relative to the first order approximation of a 2D elastic layered basin response.

## INTRODUCTION

Local site conditions such as geology, geomorphology and basin geometry have been found to have strong influence on the ground surface motion during earthquakes. Basin effects, i.e. the combined effect of entrapment and reverberation of seismic waves in soft sedimentary structures overlaying the basement rock, is one of the most important factors that has been observed in many catastrophic earthquakes such as $M_w$8.0 1985 Mexico City, Mexico, and $M_w$7.8 2015 Gorkha, Nepal. The ground surface motion which was observed in Mexico City, Mexico 1985 earthquake was a clear demonstration of basin effects: The Peak Ground Acceleration (PGA) of 0.14g in the vicinity of the epicentral area was amplified to 0.17g at a distance 200 miles from the epicenter, on the surface of clayey soft soil of the Mexico City Valley (Campillo et al (1988)). Moreover, basin effects likely dominated the ground surface motion recorded during the 2015 Gorkha earthquake sequence in Kathmandu, Nepal, as evidenced by the long duration ground motions with significantly diminished high frequency content (Rajuare et al (2016)).

Researchers have studied the basin effects in the past half a century. The pioneer work of Aki and Larner (1970), who devised a practical method to calculate the elastic wave field in a layer-over-half space medium with an irregular interface, attracted attention about this problem. Parallel to Aki and Larner, Boore (1970) studied similar problem using the Finite Difference Method (FDM) for the case of ocean-continent configuration. Later, researchers have started to build on these bases to develop analytical, semi-analytical and numerical approaches to investigate effects of soil characteristics on site response.

Initially, the in-plane (SH) wave problem attracted more attention because of its scalar nature, and, therefore, its simplicity. This followed by some researches that studied simple configurations analytically (Wong and Trifunac (1974)). For the case of more complicated geometries, others such as Sanchez Sesma and Esquivel (1979), and Bard and Bouchon (1980-I) studied basin effects for the case of out-of-plane motion numerically.

In-plane (SV-P) motions, however, were more complicated to study due to mode conversion that takes place under certain circumstances. Bard and Bouchon (1980-II) were the first who studied in-plane motion problem using the Aki-Larner method for basin structures. A few years later, Dravinski and Mossessian (1987), and Mossessian and Dravinski (1987) studied in-plane basin effects using indirect integral equation approach in one and two dipping-layer structures for elastic and damped material properties.

Later, researchers such as Kawase and Aki (1989), Sanchez Sesma et al (1993), Zhou and Chen (2008), and Martino et al (2015) used single or hybrid numerical approaches to study in-plane and out-of-plane wave propagation due to general surface irregularities and for realistic cases after earthquakes around the world.

While numerous studies have been published on the problem of basin effects so far, most of them have been performed in the frequency domain. Moreover, in most of the analyses, soil and rock materials are assumed to be elastic despite the evidence from past events that show nonlinear effects can play an important role in modifying the ground surface shaking. Therefore, in this article, we want study the basin effects observed during $M_w$7.8 2015 Gorkha earthquake by means of coupled 1D nonlinear-2D elastic model in time domain.

**METHODOLOGY**

In this article, we study 1D nonlinear and 2D elastic models of Kathmandu basin. In order to capture soil nonlinearity, we use 1D nonlinear constitutive model called HH (Shi & Asimaki (2017)) which is implemented in site response analysis code SeismoSoil. It is available online at http://asimaki.caltech.edu/resources/index.html (Last accessed September 2017). For the 2D simulation, we use an elastic 2D Finite Element Model (FEM) utilizing the Open System for Earthquake Engineering Simulation (OpenSEES) platform (Mckenna et al (2000)). Discretization is performed by quad elements and the input excitation is SV plane wave applied at the bottom of the domain. Boundary conditions will be elaborated later in this article.

**HH 1D Nonlinear Model.** Shi and Asimaki (2017) developed a nonlinear soil model, called HH, in order to capture soil response under weak and strong ground motions using just shear velocity profile as input. The HH approach is in fact a combination of two models and its formulation is as follows

$$\tau_{HH} = w(\gamma)\tau_{MKZ}(\gamma) + [1 - w(\gamma)]\tau_{FKZ}(\gamma) \tag{1}$$

in which $\tau_{MKZ}(\gamma)$ is the well-known MKZ stress (Matasovic and Vucetic (1991)), $w(\gamma)$ is a transition function, and $\tau_{FKZ}(\gamma)$ is the FKZ stress (new higher order hyperbolic model). The main advantage of this model is its ability to match both stiffness reduction and damping curves at the same time. In addition, HH follows non-Masing rules to produce more realistic hysteretic soil behavior and is able to provide satisfactory results over a broad strain range. For more detail about the model, please read Shi and Asimaki (2017).

**Free-Field Boundary Condition.** The side boundaries of our 2D model was surrounded by a special boundary condition called Free-Field (FF). The FF boundary condition, that was first implemented in the finite difference code NESSI (Cundall et al (1980)), is to impose the free-field response on the truncated boundaries. We can thereby remove the region affected by absorbing boundary conditions and reduce the total computational cost. The application of this boundary condition consists of two steps. First, prescribing appropriate dashpots called "Lysmer dashpot" parallel and perpendicular to the side boundaries (Lysmer and Kuhlemeyer (1969)).

$$C_s = \rho v_s \tag{2}$$
$$C_p = \rho v_p \tag{3}$$

in which $\rho$ is the density of the medium in neighborhood of the boundary, $v_s$ and $v_p$ are shear wave and compressional wave velocity, respectively, and, $C_s$ and $C_p$ are calculated dashpot coefficients for tangential and perpendicular directions, respectively. Second, applying the FF forces to each node on side boundaries.

$$F_x = \left(-\rho C_p(v_x^m - v_x^{ff}) - \sigma_{xx}^{ff}\right)\Delta A_y \tag{4}$$

$$F_y = \left(-\rho C_s (v_y^m - v_y^{ff}) - \sigma_{xy}^{ff}\right) \Delta A_y \tag{5}$$

$v_x^m$ and $v_y^m$ are model velocities in x and y directions, respectively. $v_x^{ff}$ and $v_y^{ff}$ are FF velocity for each direction. $\sigma_{xx}^{ff}$ and $\sigma_{xy}^{ff}$ are the FF stresses in x and y directions and $\Delta A_y$ is the element size in y direction.

Synthesizing the above, the vertical boundaries of the domain behave like 1D column. It should be noted that this method is just applicable to vertical boundaries and in the case of inclined boundaries, other types of boundary conditions should be utilized. Figure 1 shows the schematic view of the FEM domain with prescribed boundary conditions.

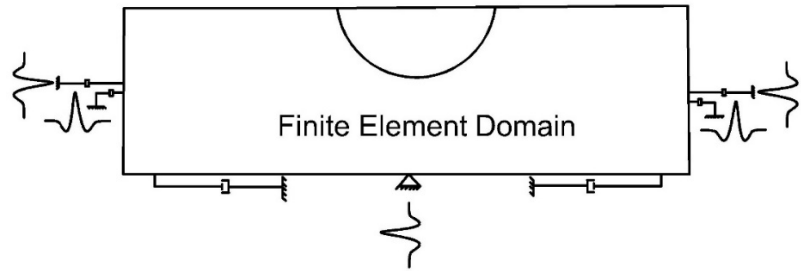

**Figure 1. Schematic view of the FEM domain and boundary conditions**

**BASIN EFFECTS DURING THE 2015 GORKHA, NEPAL EARTHQUAKE**

In this section, we show results of Kathmandu valley simulation during the Gorkha, Nepal 2015 main shock by mean of the 1D nonlinear-2D elastic models. The basin structure is non-uniform as was demonstrated by heterogeneous damage severity observed after the earthquake (Dixit et al (2015)). This causes more information to be needed which resulted in one of our main challenges in this study. This challenge includes scarcity of strong ground motion data, shear wave velocity profile and basin stratigraphy information.

In this paper, we utilize the strong ground motion data provided by Takai et al (2016). They installed four strong ground motion stations along a straight line as shown in Figure 2. One station was installed on the rock site (KTP station) and the other three were installed on the basin sediments (TVU, PTN and THM stations).

**2D Elastic Model.** Considering data scarcity, we selected a cross section of Kathmandu valley along the strong ground motion stations mentioned above (A section along black rectangle in Figure 2). Due to the lack of geological information, we utilized the basin profile information (bedrock depth at each station, velocity profile, etc) provided by Bijuckchehn et al (2016) (Figure 3-left) at strong ground motion stations and connected them using spline function to produce a continuous boundary between sedimentary layers and the bedrock (Figure 4). The general shape of basin-bedrock interface was adopted from a study accomplished by Paudyal et al (2013). They provided a 3D basin geometry using microtremor data analysis. We extracted the basin shape at the desired cross section that we are going to study in this paper and made depths consistent with information reported by Bijuckchehn et al (2016).

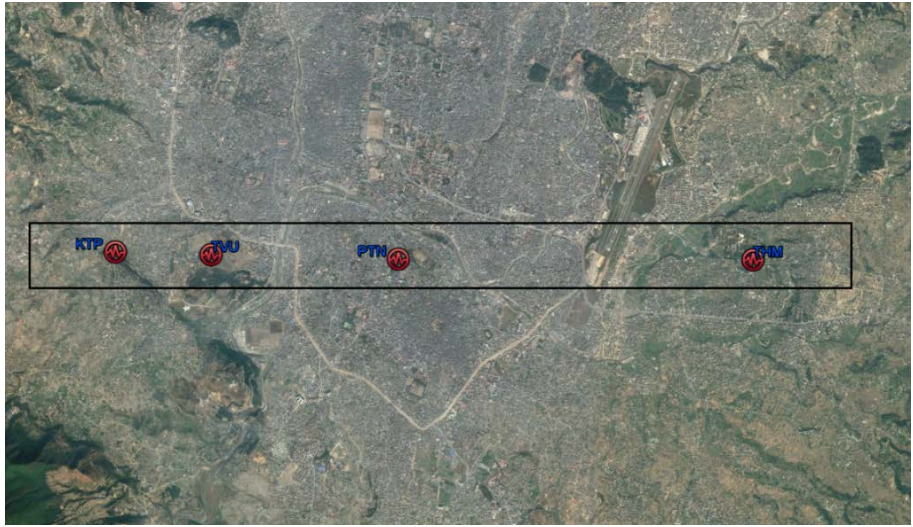

**Figure 2. Aerial view of the valley with strong ground motion stations**

The plane SV input motion was produced by means of deconvolution. For the deconvolution, a 1D column of homogeneous rock beneath the KTP station was utilized. The recorded horizontal components were rotated to the direction of SV and P waves, and, the SV component was deconvolved to produce the desired incident excitation.

Regarding material properties and shear wave velocity profile, we averaged the three available shear wave velocity profiles to the depth of 300 meters (Figure 3-right, blue line) and extrapolate the averaged data for the deeper depths up to the bedrock boundary using commonly used following equation (Figure 3-right, orange line).

$$\mathbf{V_s(z) = V_{s0} + (V_{smax} - V_{s0})(1 - e^{-kz})} \tag{6}$$

in which $V_{s0}$ and $V_{smax}$ are shear wave velocity at surface and infinite depth, respectively. k is a constant to be computed using least square fit, and, z and $V_s(z)$ are depth and shear wave velocity at the depth, respectively.

The obtained velocity profile is used in 2D simulations to prevent lateral heterogeneity. In addition, the bedrock shear wave velocity and density are assumed to be 1200 $m/s$ and 2.67 $g/cm^3$, respectively (Takai et al (2015), JICA (2002)).

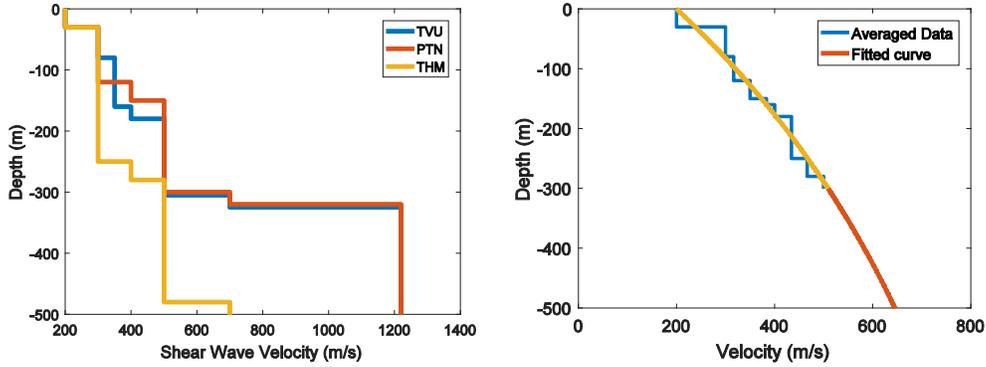

**Figure 3. Shear wave velocity profiles beneath strong ground motion stations (left) and the averaged velocity profile used in 2D simulation (right)**

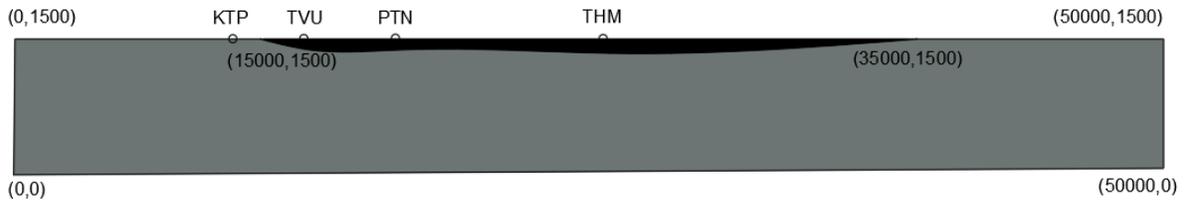

**Figure 4. Schematic view of the basin (black) and the bedrock (gray) in our 2D simulation. The four station are shown on the figure with circles. The dimensions are showing model corners and basin edges.**

Results of 2D simulations are depicted in Figure 5. We lowpass filtered all the frequencies above 1 Hz because the observed amplitude was at ~0.3 Hz. While lack of enough information was a big drawback, we could capture low-frequency content of the surface ground motion in an acceptable manner except for the TVU station. This station is the closest one to the basin edge and the discrepancy demonstrates the dominance of the basin edge effect which probably could be diminished by more sophisticated model. Moreover, the model could not behave in a realistic manner for higher frequencies due to its elastic nature, the reason why we are going to perform nonlinear analyses in next section.

**1D Nonlinear Models.** In order to have a better representation for higher frequencies, we perform nonlinear 1D simulations. In these analyzes, we use the soil profiles reported by Bijuckchehn et al (2016) (Figure 3-left) and utilize the HH constitutive model (Shi & Asimaki (2017)). The input excitation is same as 2D elastic simulations. Here, the incident motion applied as SV plane wave. As mentioned earlier, the HH model requires only shear wave velocity as input. The damping ratio and stiffness reduction curve is calculated based on the provided shear wave velocity profiles.

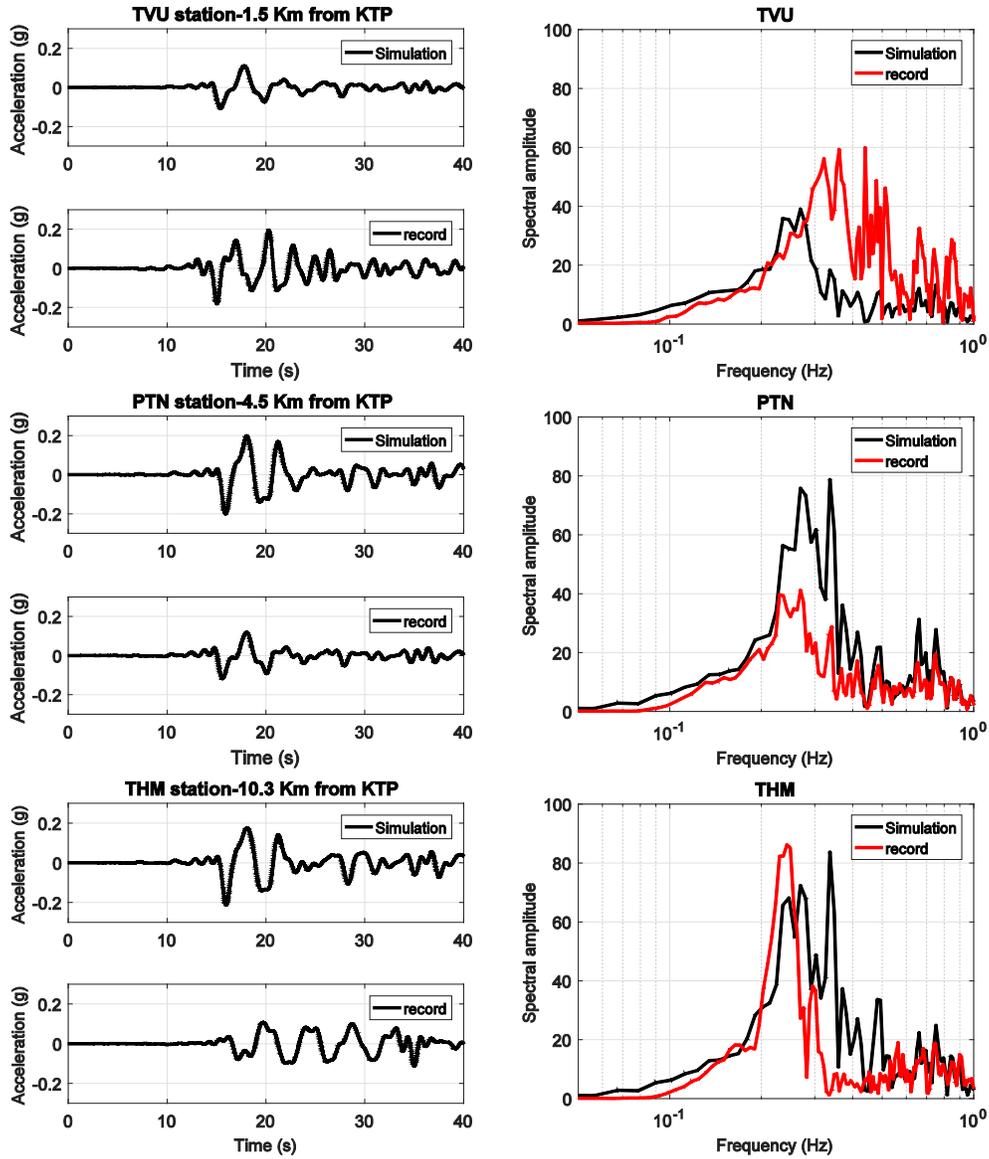
**Figure 5. Results of 2D elastic simulation, time-domain (left) and frequency domain (right)**

In order to show capability of HH constitutive model to match both stiffness reduction and damping curves simultaneously, an example is shown in Figure 6. As you can see, the match is excellent for damping curve and very good for stiffness reduction curve.

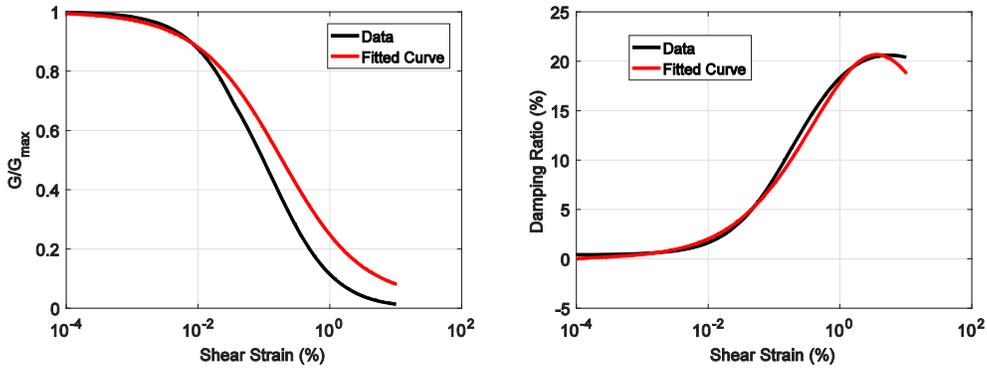

**Figure 6. An example of the HH model match. The curves are for the middle layer of TVU station**

Results of 1D analysis are low pass filtered at 3 Hz, higher frequency than elastic analysis to show applicability of nonlinear analysis to capture higher frequencies. As can be seen in Figure 7, the agreement is better than elastic model over a broader range of frequencies except for the TVU site. This shows the more complicated wave interaction happens near the basin edge as mentioned before and demonstrates the need for more appropriate model including 2D/3D geometry and a more realistic source model.

In contradiction to all the assumptions we have made for the simulations, the 1D nonlinear model confirms the possibility of the occurrence of soil nonlinearity during the main shock of the Gorkha 2015 earthquake. While we have better results for higher frequencies compared to elastic model, the model still needs to be improved in order to capture the complicated basin effects observed during the main shock of Gorkha 2015 earthquake.

**CONCLUSION**

In this article, we are trying to study the basin effects in Kathmandu valley, Nepal during the $M_w$7.8 Gorkha 2015 earthquake. A series of 1D nonlinear analyzes together with 2D elastic simulations have been carried out. Given all the simplifications of models, we were able to demonstrate the effect of nonlinearity on site response during the 1D analysis while it was obvious that a simple 1D model cannot capture the whole physics behind the basin effects and a more complicated model (i.e. 2D/3D) is required to understand the phenomenon is a more accurate way. On the other hand, the 2D elastic model is able to capture more of the basin effects in the low frequencies than high frequencies due to its elastic properties. Therefore, we are going to consider soil nonlinearity and a more realistic source model in our 2D simulations in future studies. This study was a first step towards our long term goal to develop synthetic ground motion prediction equations for the Himalayas by combining 3D nonlinear basin response simulations and kinematic source modeling.

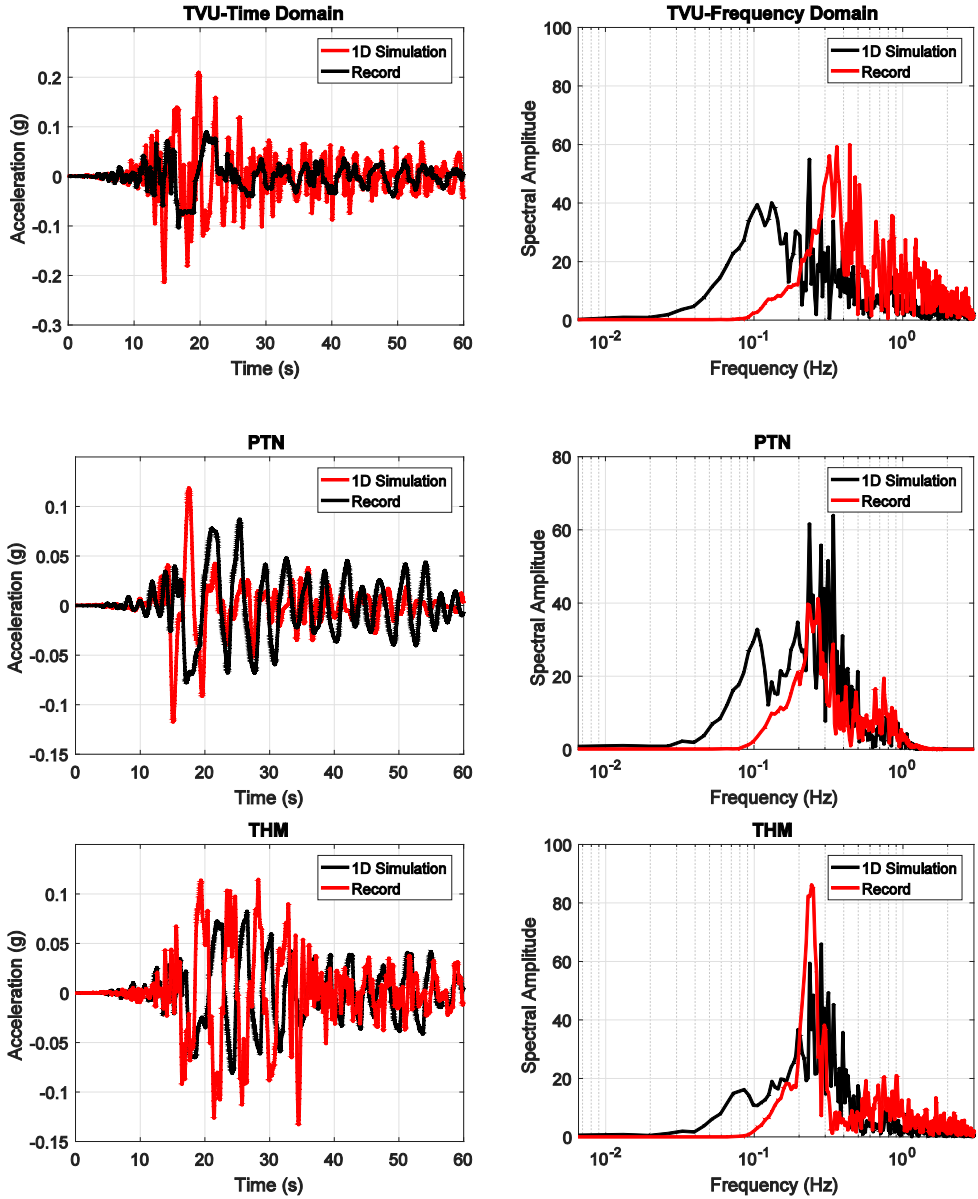

**Figure 7. Results of 1D nonlinear analysis**